\documentclass[12pt]{article}
\usepackage{epsfig}
\usepackage{graphicx}

\newcommand{\BABARPubYear}    {00}

\newcommand{\BABARProcNumber} {40}
\newcommand{\SLACPubNumber} {8731}

\newcommand\ups{$\Upsilon$(4S)}
\def\babar{\mbox{\sl B\hspace{-0.4em} {\small\sl A}\hspace{-0.4em} 
\sl B\hspace{-0.4em} {\small\sl A\hspace{-0.1em}R}}}
\def\BzBzb {\ensuremath{B^0 \Bbar^0}}
\def\Bbar  {\ensuremath{{\kern 0.2em\overline{\kern -0.2em B}}}}
\def\Bzb   {\ensuremath{\Bbar^0}}
\def\sb{\ensuremath{\sin\! 2 \beta   }}
\def\CP                 {\ensuremath{C\!P}}
\def\KS    {\ensuremath{K^0_{\scriptscriptstyle S}}}
\def\Bz    {\ensuremath{B^0}}
\def\jpsi  {\ensuremath{{J\mskip -3mu/\mskip -2mu\psi\mskip 2mu}}}
\def\psip {\ensuremath{\psi^{\prime}}}

\def\Lbabar{\mbox{{\LARGE\sl B}\hspace{-0.15em}{\Large\sl A}\hspace{-0.07em}{\LARGE\sl B}\hspace{-0.15em}{\Large\sl A\hspace{-0.02em}R}}}
\def\lbabar{\mbox{{\large\sl B}\hspace{-0.4em} {\normalsize\sl A}\hspace{-0.03em}{\large\sl B}\hspace{-0.4em} {\normalsize\sl A\hspace{-0.02em}R}}}
\usepackage{relsize}
\def\babar{\mbox{\slshape B\kern-0.1em{\smaller A}\kern-0.1em
    B\kern-0.1em{\smaller A\kern-0.2em R}}}

\def\Kbar  {\kern 0.2em\overline{\kern -0.2em K}{}}

\def\KS    {\ensuremath{K^0_{\scriptscriptstyle S}}}

\def\Kzb   {\ensuremath{\Kbar^0}}
\def\KzKzb {\ensuremath{K^0 \kern -0.16em \Kzb}}

\def\Dbar  {\kern 0.2em\overline{\kern -0.2em D}{}}

\def\Dzb   {\ensuremath{\Dbar^0}}
\def\DzDzb {\ensuremath{D^0 {\kern -0.16em \Dzb}}}

\def\Bz    {\ensuremath{B^0}}

\def\Bbar  {\kern 0.18em\overline{\kern -0.18em B}{}}

\def\Bzb   {\ensuremath{\Bbar^0}}

\def\BzBzb {\ensuremath{B^0 {\kern -0.16em \Bzb}}}

\def\jpsi  {\ensuremath{{J\mskip -3mu/\mskip -2mu\psi\mskip 2mu}}} 
\mathchardef\Upsilon="7107
\def\Y#1S{\ensuremath{\Upsilon{(#1S)}}}

\mathchardef\Deltares="7101
\mathchardef\Xi="7104
\mathchardef\Lambda="7103
\mathchardef\Sigma="7106
\mathchardef\Omega="710A
\def\Deltabar   {\kern 0.25em\overline{\kern -0.25em \Deltares}{}}
\def\Lbar {\kern 0.2em\overline{\kern -0.2em\Lambda\kern 0.05em}\kern-0.05em{}}
\def\Sigbar{\kern 0.2em\overline{\kern -0.2em \Sigma}{}}
\def\Xibar{\kern 0.2em\overline{\kern -0.2em \Xi}{}}
\def\Obar{\kern 0.2em\overline{\kern -0.2em \Omega}{}}
\def\Nbar{\kern 0.2em\overline{\kern -0.2em N}{}}
\def\Xbar{\kern 0.2em\overline{\kern -0.2em X}{}}

\def\ev   {\ensuremath{\rm \,e\kern -0.08em V}}
\def\kev  {\ensuremath{\rm \,ke\kern -0.08em V}} 
\def\mev  {\ensuremath{\rm \,Me\kern -0.08em V}} 
\def\gev  {\ensuremath{\rm \,Ge\kern -0.08em V}} 
\def\gevc {\ensuremath{{\rm \,Ge\kern -0.08em V\!/}c}} 
\def\tev  {\ensuremath{\rm \,Te\kern -0.08em V}}
\def\mevc {\ensuremath{{\rm \,Me\kern -0.08em V\!/}c}} 
\def\gevcc{\ensuremath{{\rm \,Ge\kern -0.08em V\!/}c^2}} 
\def\mevcc{\ensuremath{{\rm \,Me\kern -0.08em V\!/}c^2}}

\def\mus  {\ensuremath{\rm \,\mus}}

\def\mus        {\ensuremath{\,\mu{\rm s}}}    
\def\gsim{{~\raise.15em\hbox{$>$}\kern-.85em
          \lower.35em\hbox{$\sim$}~}}
\def\lsim{{~\raise.15em\hbox{$<$}\kern-.85em
          \lower.35em\hbox{$\sim$}~}}

\def\CP                 {\ensuremath{C\!P}}

\def\to                 {\ensuremath{\rightarrow}}

\def\pep2{PEP-II}

\def\sb{${\sin\! 2 \beta   }$}

\providecommand{\eqref}[1]{Eq.~(\ref{eq:#1})}

\def\jetset74   {\mbox{\tt Jetset \hspace{-0.5em}7.\hspace{-0.2em}4}}

\long\def\inst#1{\par\nobreak\kern 4pt\nobreak
    {\it #1}\par\vskip 10pt plus 3pt minus 3pt}

\begin{document}
{\pagestyle{empty}

\begin{flushright}
SLAC-PUB-\SLACPubNumber \\
\babar-PROC-\BABARPubYear/\BABARProcNumber \\
December, 2000 \\
\end{flushright}

\par\vskip 4cm

\begin{center}
\Large \bf The \Lbabar\ Measurement of \sb\ and its Future Prospects
\end{center}
\bigskip

\begin{center}
\large 
James Weatherall \\
University of Manchester \\
Department of Physics and Astronomy, University of Manchester, \\ 
Oxford Road, Manchester, M13 9PL, UK \\
(for the \lbabar\ Collaboration)
\end{center}
\bigskip \bigskip

\begin{center}
\large {\bf Abstract} \\ \bigskip
The measurement of \sb\ by the \babar\ experiment, where $\beta$ is one of 
the angles of the Unitarity Triangle, is described.  Some prospects for the 
future of the measurement are also discussed.
\end{center}

\vfill
\begin{center}
Contributed to the Proceedings of the UK Phenomenology Workshop \\ 
on Heavy Flavour and \CP\ Violation, \\
9/17/2000---9/22/2000, Durham, UK
\end{center}

\vspace{1.0cm}
\begin{center}
{\em Stanford Linear Accelerator Center, Stanford University, 
Stanford, CA 94309} \\ \vspace{0.1cm}\hrule\vspace{0.1cm}
Work supported in part by Department of Energy contract DE-AC03-76SF00515.
\end{center}

\section{Introduction}
The \babar\ experiment consists of an asymmetric electron-positron collider
operating at the \ups\ resonance.  More details on the detector can be found
in~\cite{Verderi:2000pv}.  The aim is to overconstrain the
unitarity triangle by measuring its sides and angles.  The analysis reviewed
here measures \sb\ by studying time-dependent \CP\ violating asymmetries in
$\Bz \rightarrow \jpsi \KS$ and $\Bz \rightarrow \psip \KS$ decays.

\section{Overview of the \sb\ analysis}
There are five main parts to measuring the \CP\ violating asymmetry:

\begin{itemize}
\item Selection of signal \CP\ events
\item Measurement of the distance ${\rm \Delta} z$ between the two \Bz\ decay 
vertices along the \ups\ boost axis
\item Determination of the flavour of the tag-side B
\item Measurement of dilution factors for the different tagging categories
\item Extraction of \sb\ via an unbinned maximum likelihood fit
\end{itemize}

\subsection{Event Selection}
The sample used for the analysis is 9.8 fb$^{-1}$ of data recorded between 
January and July 2000 of
which 0.8 fb$^{-1}$ was recorded 40 MeV below the \ups\ resonance.  Particle
identification uses mainly the CsI calorimeter for electrons, the Instrumented
Flux Return for muons and the DIRC for kaons.  Extra information is provided 
by dE/dx measured in the tracking system.  
The selection for the \CP\ events proceeds as follows.
Pairs of electrons or muons coming
from a common vertex are combined to form \jpsi\ and \psip\ candidates.  
The \psip\ is also reconstructed from its decay into \jpsi\ $\pi^+\pi^-$.
The $K_S$ candidates are made from either a pair of charged tracks or a pair
of $\pi^0$ candidates.  In addition there are various event shape and 
topological cuts designed to reduce continuum and $B\overline{B}$ background.
Full details of the selection can be found in~\cite{Hitlin:2000tm}.
The final event sample is shown in figure~\ref{fig:sigsam}.

There are two other $B$ decay samples.  One consists of fully reconstructed
semileptonic ($B^0 \to D^{*-}l^+\nu_l$) and hadronic 
($B^0 \to D^{(*)-}\pi^+,D^{(*)-}\rho^+,D^{(*)-}a_1^+$) decays as well as a
control sample of $B^+ \to \overline{D}^{(*)0}\pi^+$ events.  The selection
of this sample is described in~\cite{Aubert:2000sz} and~\cite{Aubert:2000vt}.
The other is a charmonium control sample containing fully reconstructed 
neutral or charged $B$ candidates in two-body decay modes with a \jpsi\ in
the final state (e.g. $B^+ \to \jpsi\ K^+, B^0 \to \jpsi\ K^{*0}
(K^{*0} \to K^+\pi^-)$).

\begin{figure}[htbp]
\begin{center}
\epsfig{file=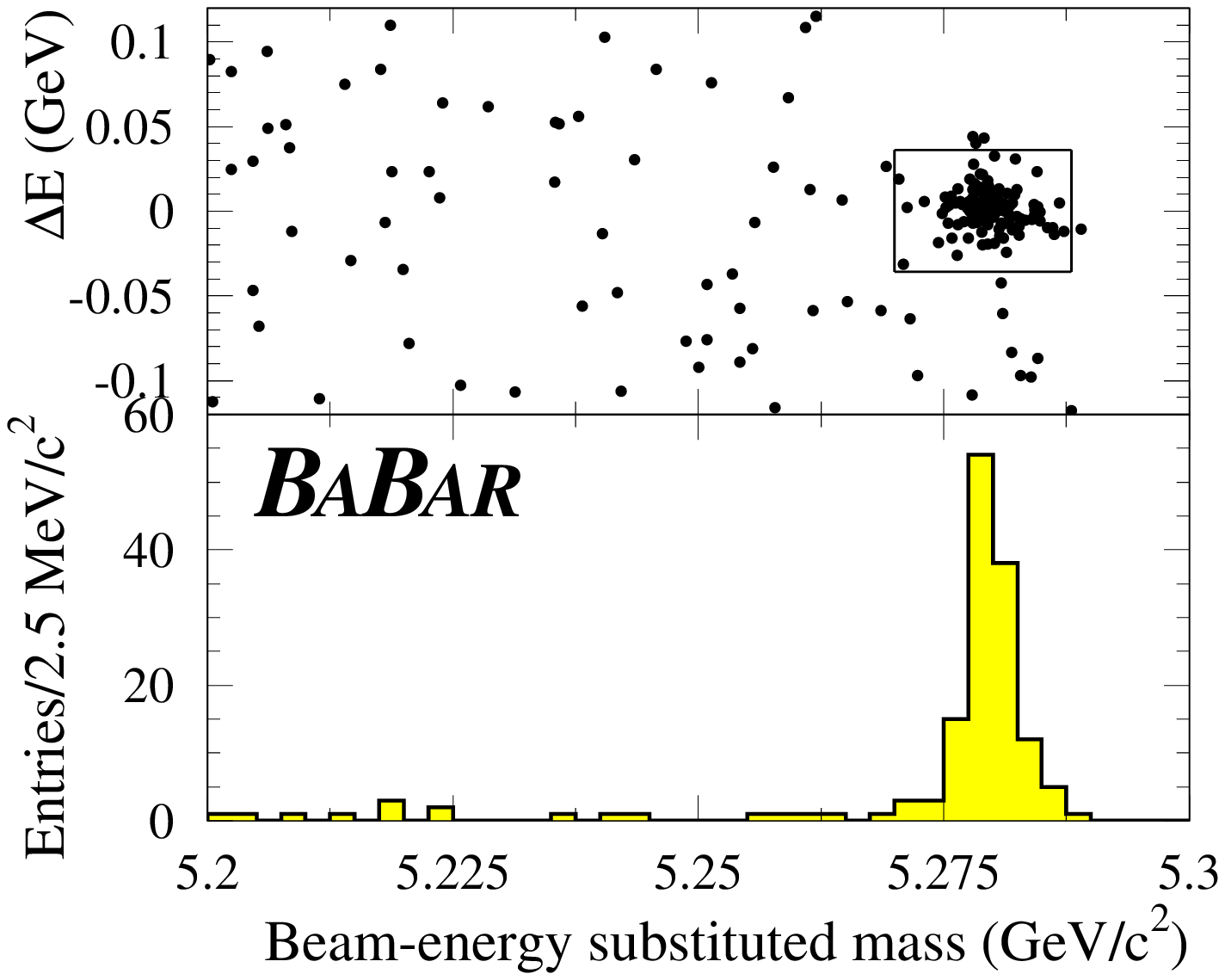,width=5cm}
\epsfig{file=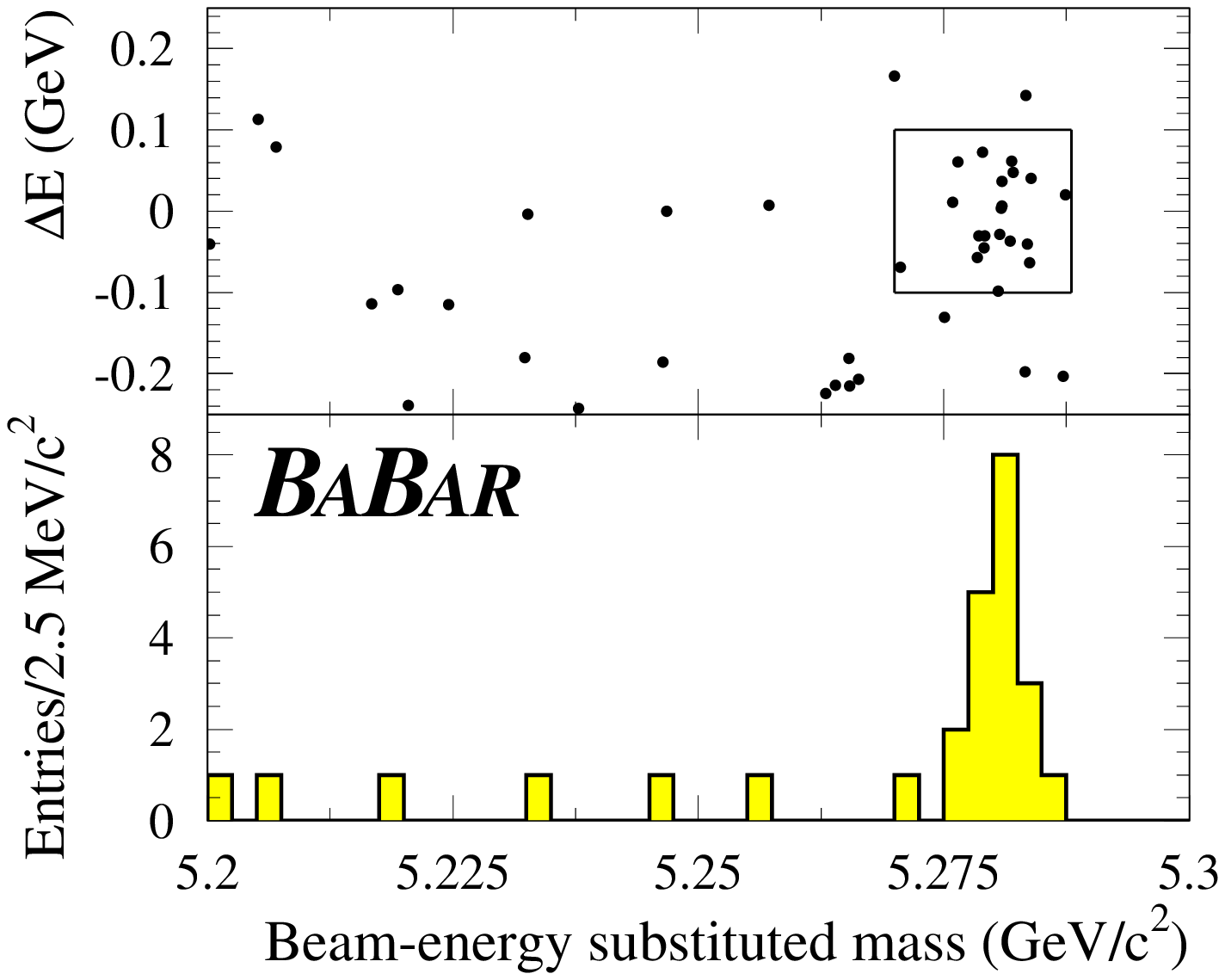,width=5cm}
\epsfig{file=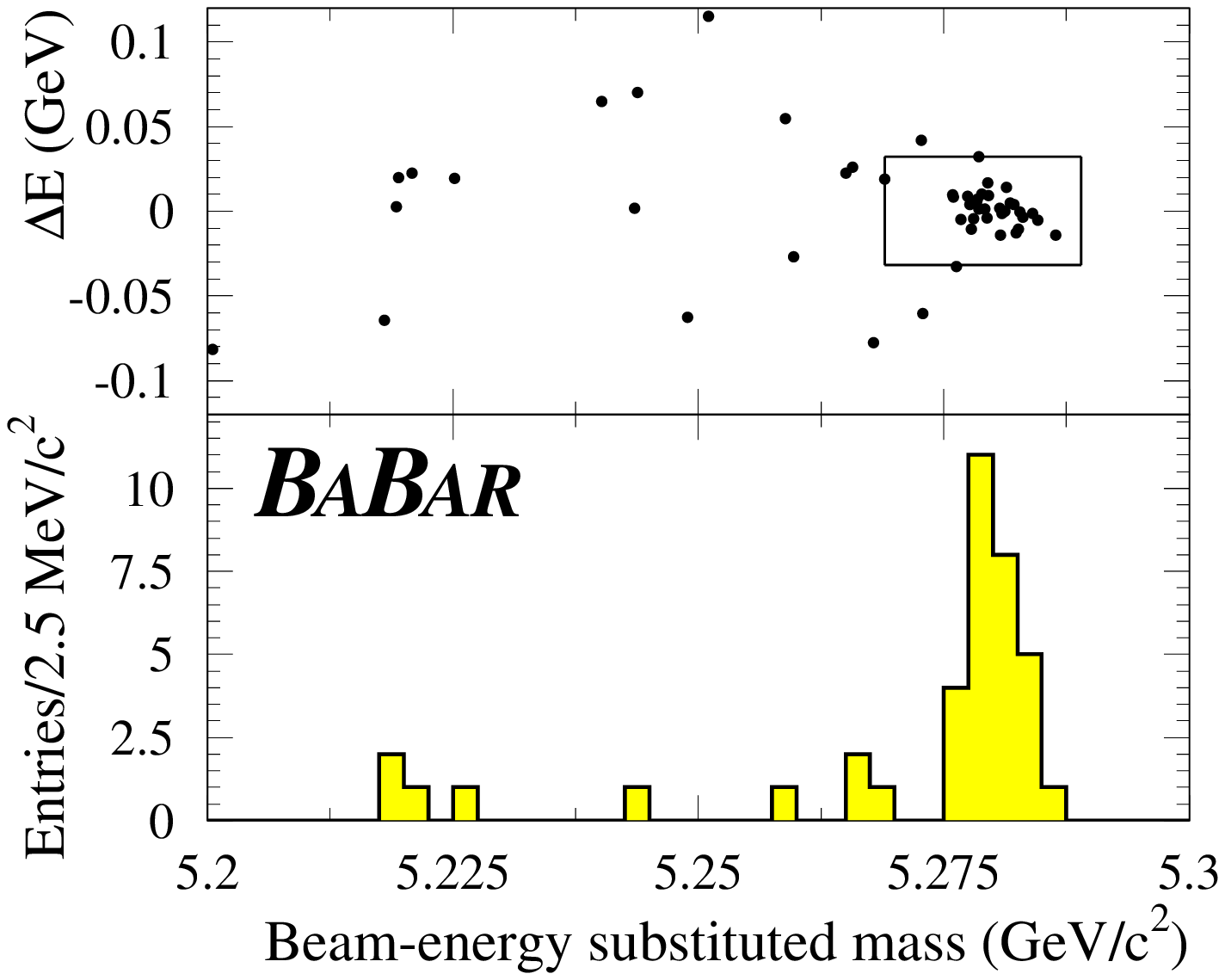,width=5cm}
\end{center}
\caption{\CP\ signal event distributions for $J/\psi K_S(\pi^+\pi^-)$ 
(left), $J/\psi K_S(\pi^0\pi^0)$ (middle) and $\psip\ K_S (\pi^+\pi^-)$ 
(right).}
\label{fig:sigsam}
\end{figure}

\subsection{Measuring ${\rm \Delta} z$}
The time-dependent decay rate for the $B_{\CP}$ is given by
\begin{equation}
\label{eq:TimeDep}
f_\pm(\, \Gamma, \, {\rm \Delta} m_d, \, {\cal {D}} \sin{ 2 \beta }, \, 
t \, )  = {\frac{1}{4}}\, \Gamma \, {\rm e}^{ - \Gamma \left| t \right| }
\, \left[  \, 1 \, \pm \, {\cal {D}} \sin{ 2 \beta } \times \sin{ {\rm \Delta}
m_d \, t } \,  \right]
\end{equation}
where the + or - sign indicates whether the $B_{tag}$ was tagged as a $B^0$
or $\Bzb$ respectively.  The dilution factor $\cal{D}$ is given by
${\cal D}=1-2w$, where $w$ is the mistag fraction (the probability that the 
$B_{tag}$ is identified incorrectly).  To account for finite detector 
resolution, the time distribution must be convoluted with a resolution 
function:

\begin{eqnarray}
{\cal {R}}( {\rm \Delta}z ; \hat {a} \,  ) &=& \sum^{2}_{i=1} \, \frac{f_i}
{\sigma_i\sqrt{2\pi}} \, {\rm exp} \left(  - ( {\rm \Delta}z-\delta_i)^2/2
{\sigma_i}^2   \right) \ \ \ , 
\end{eqnarray}

\noindent
which is just the sum of two Gaussians where the $f_i$, $\delta_i$ and 
$\sigma_i$ are the normalizations, biases and widths of the distributions.
In practice two scale factors ${\cal S}_1$ and ${\cal S}_2$ are introduced 
such that $\sigma_i={\cal S}_i\times\sigma_{\Delta t}$ where 
$\sigma_{\Delta t}$ is an event-by-event calculated error on $\Delta t$.
They take account of underestimating the uncertainty on $\Delta t$
due to effects such as hard scattering
and possible underestimation of the amount of material traversed by the 
particles.
The resolution function parameters are obtained from a maximum likelihood 
fit to the hadronic $B^0$ sample and are shown in table~\ref{tab:res}.  The 
$f_w$ parameter represents the width of a third Gaussian component, included
to accommodate a small ($\sim$1\%) 
fraction of events which have very large values 
of $\Delta z$, mostly caused by vertex reconstruction problems.  This 
Gaussian is unbiased with a fixed width of 8 ps.
Further details can be found in~\cite{Aubert:2000sz}.

\begin{table}
\vspace{0.3cm}
\begin{center}
\begin{tabular}{|cc|cc|} \hline
\multicolumn{2}{|c|}{Parameter} & \multicolumn{2}{c|}{Value} \\ \hline \hline
$\delta_1$   & (ps) & $-0.20\pm0.06$ & from fit \\ 
${\cal S}_1$ &      & $1.33\pm0.14$  & from fit \\ 
$f_w$        & (\%) & $1.6\pm0.6$ & from fit \\ 
$f_1$        & (\%) & $75$ & fixed \\ 
$\delta_2$   & (ps) & $0$ & fixed \\ 
${\cal S}_2$ &      & $2.1$ & fixed \\ 
\hline
\end{tabular}
\end{center}
\caption{Resolution function parameters.  Those, labeled 'from fit' are 
measured from data and those marked 'fixed' are determined from Monte Carlo.} 
\label{tab:res}
\end{table}

\subsection{B flavour tagging}
Each event with a \CP\ candidate is assigned a $B^0$ or $\Bzb$ tag 
if it satisfies the criteria for one of the several tagging categories.  The
figure of merit for each tagging category is the effective tagging efficiency
$Q_i = \epsilon_i(1-2w_i)^2$ where $\epsilon_i$ is the fraction of events 
assigned to category $i$ and $w_i$ is the probability of mis-tagging an event
in this category.  The statistical error on \sb\ is proportional to 
$1/\sqrt{Q}$ where $Q=\sum_iQ_i$.  
There are five tagging categories: {\tt Electron, Muon,
Kaon, NT1 and NT2}.  

The first three require the presence of a fast lepton 
and/or one or more charged kaons in the event and depend on the correlation 
between the charge of a primary lepton or kaon and the flavour of the $b$ 
quark.  If an event is not assigned to either the {\tt Electron} or {\tt Muon}
categories, it is assigned to the {\tt Kaon} category if the sum of the 
charges of all the identified kaons in the event is different from zero.  If
both lepton and kaon tags are available but inconsistent the event is rejected
from both categories.  

NT1 and NT2 are categories from a neural network algorithm, 
this approach being motivated by 
the potential flavour-tagging power carried by non-identified leptons and 
kaons, correlations between leptons and kaons and more generally the momentum
spectrum of charged particles from $B$ meson decays.  The output of the neural
network tagger $x_{NT}$ can be mapped onto the interval [-1,1] with 
$x_{NT}<0$ representing a $B^0$ tag and $x_{NT}>0$ a $\Bzb$ tag.  Events with
$|x_{NT}|>0.5$ are classified in the {\tt NT1} category and events with 
$0.2<|x_{NT}|<0.5$ in the {\tt NT2} category.  Events with $|x_{NT}|<0.2$ are
excluded from the final analysis sample.

\subsection{Measurement of tagging performance}
The effective tagging efficiencies and mistag fractions
for all the categories are measured from data using a maximum likelihood fit
to the time distributions of the $B^0$ hadronic event sample.  The procedure
uses events which have one $B$ fully reconstructed in a flavour eigenstate 
mode.  The tagging algorithms are then applied to the rest of the event, which
represents the potential $B_{tag}$.  Events are classified as {\em mixed} or
{\em unmixed} depending on whether the $B_{tag}$ is tagged with the same or
opposite flavour as the $B_{CP}$.  One can express the time-integrated 
fraction of mixed events $\chi$ as a function of the \BzBzb\ mixing 
probability, $\chi = \chi_d+(1-2\chi_d)w$
where $\chi_d=\frac{1}{2}x^2_d/(1+x^2_d)$, with $x_d=\Delta m_d/\Gamma$.  Thus
an experimental value of the mistag fraction $w$ can be deduced from the data.

A more accurate estimate of $w$ comes from a time-dependent analysis of the 
fraction of mixed events.  The mixing probability is smallest at low 
$\Delta t$ so that this region is governed by the mistag fraction.  
Figure~\ref{fig:mix} shows the fraction of mixed events versus $\Delta t$.
The resultant tagging performances are shown in table~\ref{tab:tag}.  

\begin{figure}[htbp]
\begin{center}
\epsfig{file=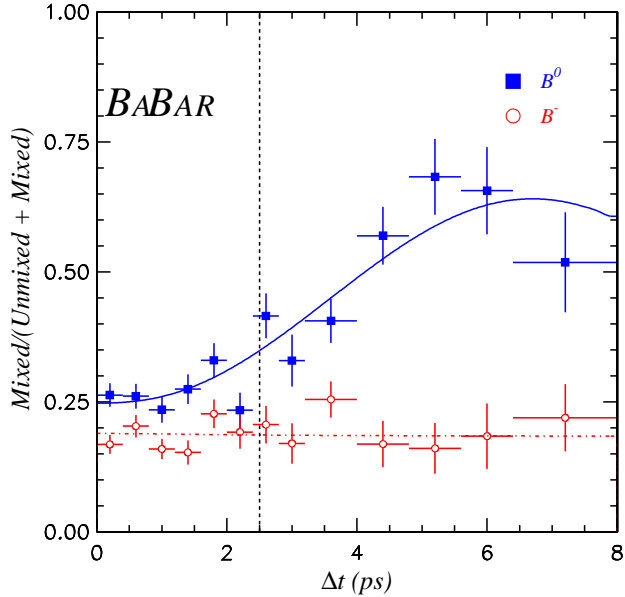,width=9cm}
\end{center}
\caption{The fraction of mixed events as a function of $|\Delta t|$ for
data events in the hadronic sample for neutral $B$ mesons (full squares) and
charged $B$ mesons (open circles).  The dot-dashed line at $t_{cut}=2.5$ ps
indicates the bin boundary for the time-integrated single-bin method.}
\label{fig:mix}
\end{figure}

\begin{table}
\vspace{0.3cm}
\begin{center}
\begin{tabular}{|l|c|c|c|} \hline
Tagging category & $\epsilon$ (\%) & $w$ (\%) & $Q$ (\%) \\ \hline \hline
{\tt Lepton}  & $11.2\pm0.5$ & $9.6\pm1.7\pm1.3$  & $7.3\pm0.3$  \\
{\tt Kaon}    & $36.7\pm0.9$ & $19.7\pm1.3\pm1.1$ & $13.5\pm0.3$ \\
{\tt NT1}     & $11.7\pm0.5$ & $16.7\pm2.2\pm2.0$ & $5.2\pm0.2$  \\
{\tt NT2}     & $16.6\pm0.6$ & $33.1\pm2.1\pm2.1$ & $1.9\pm0.1$  \\ \hline
\hline
all           & $76.7\pm0.5$ &                    & $27.9\pm0.5$  \\
\hline
\end{tabular}
\end{center}
\caption{Tagging performance as measured from data.} 
\label{tab:tag}
\end{table}

\subsection{Extracting \sb}
A blind analysis technique was adopted for the extraction of \sb\ to 
eliminate possible experimenter bias.  The technique hides both the result of
the likelihood fit and the visual \CP\ asymmetry in the $\Delta t$
distribution.  This method allows systematic studies to be performed while 
keeping the numerical value of \sb\ hidden.

Possible systematic effects due to uncertainty in the input parameters to the
fit, incomplete knowledge of the time resolution function, uncertainties in the
mistag fractions and possible limitations in the analysis procedure were all
studied.  Details can be found in~\cite{Hitlin:2000tm}.  The systematic 
errors are summarized in table~\ref{tab:syst}.

\begin{table}
\vspace{0.3cm}
\begin{center}
\begin{tabular}{|l|c|} \hline 
Source of uncertainty & Uncertainty on \sb\ \\ \hline \hline
uncertainty on $\tau^0_B$ & 0.002 \\
uncertainty on $\Delta m_d$ & 0.015 \\
uncertainty on $\Delta z$ resolution for \CP\ sample & 0.019 \\
uncertainty on time-resolution bias for \CP\ sample & 0.047 \\
uncertainty on measurement of mistag fractions & 0.053 \\
different mistag fractions for \CP\ and non-\CP\ samples & 0.050 \\
different mistag fractions for \Bz\ and \Bzb\ & 0.005 \\
background in \CP\ sample & 0.015 \\ \hline \hline
total systematic error & {\bf 0.091} \\
\hline
\end{tabular}
\end{center}
\caption{Summary of systematic uncertainties.  The different contributions are
added in quadrature.}
\label{tab:syst}
\end{table}

\subsection{Results and checks}
The maximum likelihood fit for \sb, using the full tagged sample of 120
$\Bz \rightarrow \jpsi \KS$ and $\Bz \rightarrow \psip \KS$ events yields:

\begin{equation}
\sin\! 2 \beta = 0.12 \pm 0.37\, (stat) \pm 0.09\, (syst)\ (preliminary).
\end{equation}

\noindent
The log likelihood is shown as a function of \sb\ in figure~\ref{fig:like}.
The raw asymmetry as a function of $\Delta t$ is shown in figure~\ref{fig:raw}

\begin{figure}
\begin{center}
\epsfig{file=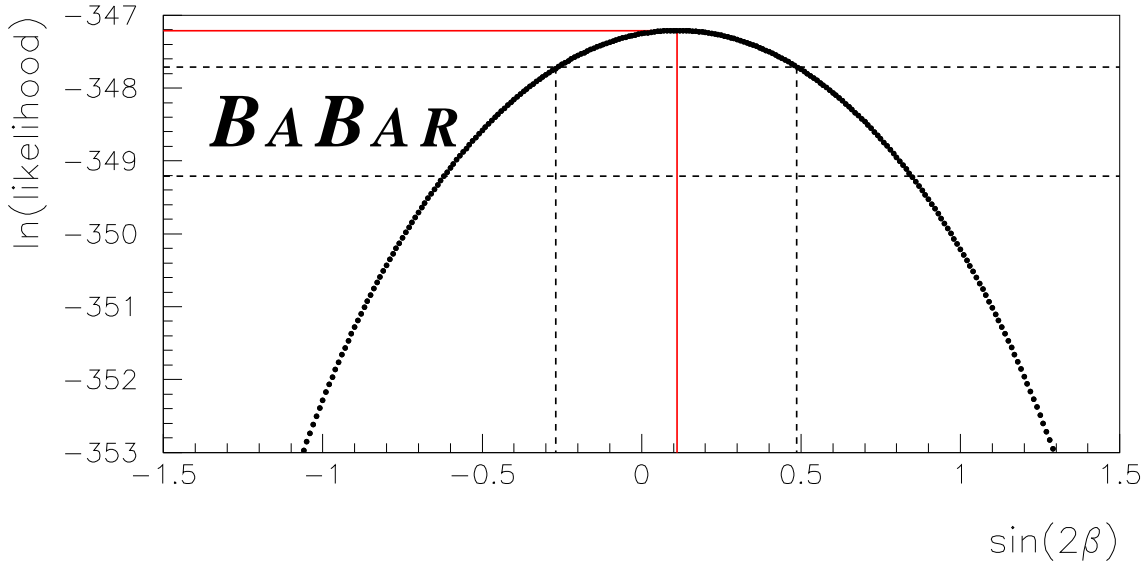,height=9cm}
\caption{Variation of the log likelihood as a function of \sb.  The 
two horizontal dashed lines indicate changes in the log-likelihood 
corresponding to one and two statistical standard deviations.}
\label{fig:like}
\end{center}
\end{figure}

\begin{figure}
\begin{center}
\epsfig{file=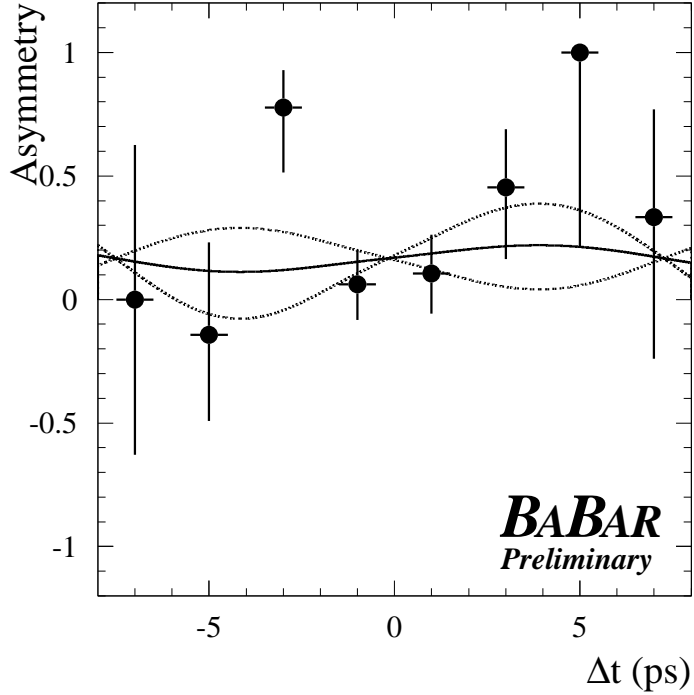,width=9cm}
\caption{The raw $B^0-\Bzb$ asymmetry $(N_{B^0}-N_{\Bzb})/
(N_{B^0}+N_{\Bzb})$.  The time-dependent asymmetry is represented by a solid
curve for the central value of \sb, and by two dotted curves for the values at
plus and minus one statistical standard deviation from the central value.  The
curves are not centered at (0,0) because the \CP\ sample contains an unequal 
number of \Bz\ and \Bzb\ events (70 \Bz\ versus 50 \Bzb).  The $\chi^2$ 
between the binned asymmetry and the result of the maximum likelihood fit is
9.2 for 7 degrees of freedom.}
\label{fig:raw}
\end{center}
\end{figure}

The probability of obtaining a statistical uncertainty of 0.37 is estimated
by generating a large number of toy Monte Carlo experiments with the same 
number of tagged \CP\ events as in the data sample.  The errors are
distributed around 0.32 with a standard deviation of 0.03, meaning that the
probability of obtaining a larger statistical error that the one observed is
5\%.  From a large number of full Monte Carlo simulated experiments, we 
estimate that the probability of finding a lower value of the likelihood than
the one observed is 20\%.

Several cross-checks are performed to validate the main analysis.  The 
charmonium and fully-reconstructed hadronic control samples are composed of 
events that should exhibit no time-dependent asymmetry.  These events are 
fitted in the same way as the signal \CP\ events to extract an ``apparent
\CP\ asymmetry''.  The results are shown in table~\ref{tab:chk}.  

\begin{table}
\vspace{0.3cm}
\begin{center}
\begin{tabular}{|l|c|} \hline 
Sample & Apparent \CP\ asymmetry \\ \hline \hline
Hadronic charged $B$ decays & $0.03\pm0.07$ \\ \hline
Hadronic neutral $B$ decays & $-0.01\pm0.08$ \\ \hline
$J/\psi K^+$ & $0.13\pm0.14$ \\ \hline
$J/\psi K^{*0} (K^{*0} \to K^+\pi^-)$ & $0.49\pm0.26$ \\ \hline
\end{tabular}
\end{center}
\caption{Summary of systematic uncertainties.  The different contributions are
added in quadrature.}
\label{tab:chk}
\end{table}    

\subsection{Constraints on the Unitarity Triangle}
The Unitarity Triangle in the ($\overline{\rho},\overline{\eta}$) plane is 
shown in figure~\ref{fig:ut}.  The two solutions corresponding to the 
measured central value are shown as 
straight lines.  The cross-hatched regions correspond to one and two times 
the one-standard-deviation experimental uncertainty.  The ellipses represent
regions allowed by all other measurements that constrain the triangle.  They
are shown for a variety of choices of theoretical parameters.  More details
can be found in~\cite{Harrison:1998yr}.

\begin{figure}
\begin{center}
\epsfig{file=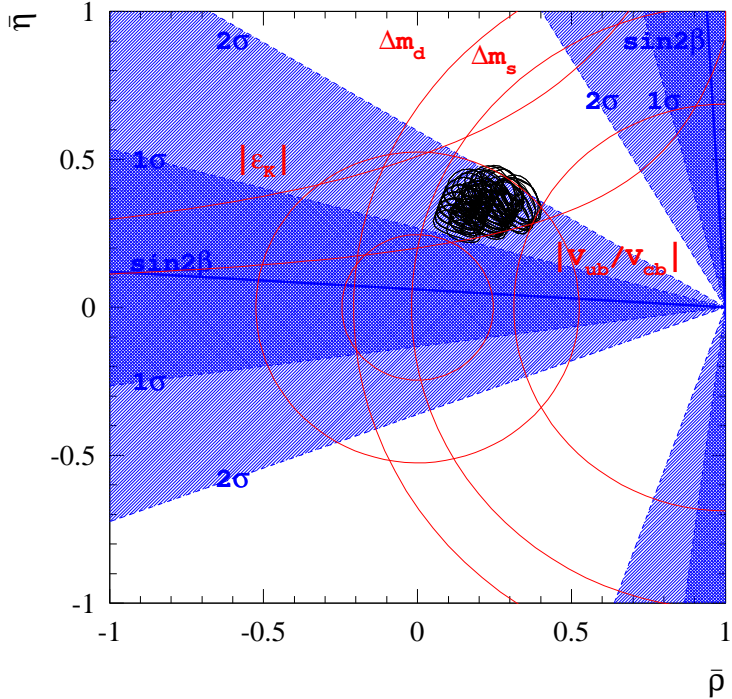,width=10cm}
\caption{Present constraints on the position of the apex of the Unitarity
Triangle with the \babar\ result indicated by the cross-hatched regions.}
\label{fig:ut}
\end{center}
\end{figure}

\section{Future prospects}
The preceding pages describe only a preliminary measurement of \sb\ by 
the \babar\ experiment.  More data will allow extra channels to be included in
the final fit as well as providing more events for the currently used decay 
modes.  The new channels will bring extra experimental and theoretical 
challenges with them.  Such present and future issues are discussed in the 
sections that follow.

\subsection{Available Modes}
The B decay modes that have been used to measure \sb\ up to now are clean
in that they are vector-scalar, $b \to ccs$ transitions which have no 
significant pollution from penguin diagrams.  The next step is to add 
vector-vector modes such as $B^0 \to J/\psi K^*$.  These modes require an 
angular analysis of the
vector meson decay products, due to the different partial waves and 
therefore admixture of \CP\ odd and \CP\ even that is present in the final state.
Such an angular analysis has already yielded preliminary results for the 
$J/\psi K^*$ modes.  Once one has measured the polarizations in these modes, 
they are as clean, theoretically, as the vector-scalar modes.
Another obvious addition is $B^0 \to J/\psi K_L$ decays where the challenge
here is to understand the background well enough to make the channel feasible.
Work is ongoing in this area.

A different kind of difficulty is presented by channels with a significant
degree of penguin contamination, such as $b \to ccd$ scalar-scalar modes
(e.g. $B^0 \to D^+ D^-$).  Here the fit must take into account the fact that
the true value of \sb\ is shifted by an amount proportional to the ratio
of tree to penguin contributions.  This ratio is model dependent and subject
to large theoretical uncertainties.

Finally, modes such as $B^0 \to D^* D^*$ and $B \to J/\psi \rho^0$ which are
vector-vector, $b \to ccd$ transitions face the theoretical challenges of the
penguin contaminated modes described above, as well as requiring an angular
analysis to solve the vector-vector \CP\ admixture problem.

\subsection{Experimental Considerations}
There are also experimental analysis issues which need to be resolved or 
studied in greater
depth in the future.  The tagging algorithms that \babar\ uses should be
developed and extended to include extra tagging categories such as the using
the soft pion from $D^*$ decays and incorporating leptons at an intermediate 
momentum (i.e. from a cascade).  It would be useful to take account of 
correlations within an event, such as when two different tagging categories 
report an answer.  This can give more information about the event if the 
correlations are well understood.  There is also an open question when it 
comes to measuring the tagging performance from the hadronic or semileptonic 
$B$ decay samples.  One then needs to be absolutely sure that using exactly
the same numbers for the \CP\ signal event sample is a valid thing to do.

The measurement of $\Delta z$ is another crucial part of the analysis and it is
important that the errors and biases to this distribution are understood.
The distribution tends to be biased by the decays of particles which fly 
significantly 
from the original $B$ decay vertex, such as $D^0$s.  These can be rejected
by looking explicitly for cascade decays.  The parameterization of the 
resolution function incorporates detector effects such as misalignments and 
electronics readout effects.  All contributions to the width should be 
studied in order to fully understand the error on $\Delta z$.

Backgrounds to the various \CP\ modes can also be a problem.  The channels vary
in terms of how much background they experience and this background can be
particularly dangerous if it has a significant structure in $\Delta z$.
For charmonium channels, much of the background comes from events containing
a real $J/\psi$.  In that case, one needs to study exactly which modes 
contribute and what their shape is in the final distributions (if they cannot
be removed otherwise).  Non-resonant backgrounds to vector-vector modes such
as the $J/\psi K^0 \pi^0$ contribution to $J/\psi K^{*0}(K^{*0} \to K^0\pi^0)$
are in principle dangerous since they can have \CP\
violating properties but no angular structure.  However, the branching ratios 
for these non-resonant modes are typically poorly known and consistent with 
zero making it difficult to simulate them in the correct proportions.

\subsection{Study of Statistical Error}
It seems anomalous that both \babar\ and Belle record higher statistical errors
than one would expect.  The fitting procedure is, and continues to be a
vigorously studied part of the analysis as we need to be certain that the
likelihood function is of exactly the correct form for the final fit.

\section{Conclusions}
A preliminary measurement of \sb\ by \babar\ has been presented.  The errors
on the final result make it difficult to express its significance in terms
of constraints on the Unitarity Triangle.  However, results based on a much 
larger data sample ($\sim$20 fb$^{-1}$) will soon be available.  Combined
with a better understanding of systematic effects, this should make the next
measurement of \sb\ even more interesting than the current one.  It is also
expected that other \CP\ modes will soon be available for
analysis including $B^0 \to J/\psi K^{*0} (K^{*0} \to K_S \pi^0)$ and
$B^0 \to J/\psi K_L$.  The larger data sample with additional \CP\ modes 
should yield a value of \sb\ for which the statistical and systematic errors
are about one-half of their current values.


\end{document}